\documentstyle[epsf]{mn}

\newcommand{\lya}{Ly$\alpha$\ }
\title[GMRT Observations of Low z Damped~\lya Absorbers]
  {GMRT Observations of Low z Damped~\lya Absorbers}
\author[Jayaram N. Chengalur \& Nissim Kanekar]
  {Jayaram N. Chengalur$^1$\thanks{chengalur@ncra.tifr.res.in}
   and Nissim Kanekar$^1$\thanks{nissim@ncra.tifr.res.in} \\
  $^1$National Center for Radio Astrophysics (TIFR),
	University of Pune Campus, P. O. Bag 3, Ganeshkhind, Pune 411007,
	India.\\}
\date{}
\pagerange{\pageref{firstpage}--\pageref{lastpage}}

\def\LaTeX{L\kern-.36em\raise.3ex\hbox{a}\kern-.15em
    T\kern-.1667em\lower.7ex\hbox{E}\kern-.125emX}

\begin{document}

\label{firstpage}

\maketitle

\begin{abstract}

	We present Giant Metrewave Radio Telescope (GMRT) observations 
of redshifted HI 21cm absorption in two low redshift ($z=0.2212,~z=0.0912$)
damped~\lya systems seen towards the Gigahertz peaked source 
OI~363 ($z_{em} = 0.630$). The object at $z=0.0912$ is the lowest redshift 
damped~\lya system known to date. Ground based imaging \cite{rt} shows
that at neither redshift is there a large spiral galaxy at low impact
parameter to the line  of sight to OI~363, in contradiction with the 
suggestion that damped~\lya systems are large proto-disks. Since OI~363 is
a highly compact, core dominated source, the covering factor of the HI gas 
is likely to be unity. 
Nonetheless, the spin temperatures derived from the 21cm optical depth 
(and using the $N_{HI}$ measured from HST spectra \cite{rt}) are high, 
viz.  $1120 \pm 200$~K and $825 \pm 110$~K for the high and low redshift 
systems respectively. These values are considerably higher than typical 
values ($100 - 200$~K) measured in our Galaxy and Andromeda and are, 
in fact, similar to those obtained in high redshift
damped~\lya systems. Our observations hence suggest that evolutionary
effects may not be crucial in understanding the difference in derived spin 
temperature values between local spiral disks and high redshift 
damped~\lya systems.

\end{abstract}


\section{Introduction}

	Damped~\lya systems are the extremely high HI column density 
($N_{HI} > 2\times 10^{20}$ atoms~cm$^{-2}$) systems seen in absorption in
the spectra taken towards distant quasars. Although rare, they are the 
dominant contributors (by mass) to the observed neutral gas at high 
($z \sim 3$) redshifts. Principally for this reason,  these systems are 
natural candidates for the precursors of $z=0$  galaxies. Consistent
with this interpretation, the mass density of neutral gas in damped~\lya
systems at  $z \sim 3$ is comparable to the mass density in stars in 
luminous galaxies at $z=0$. Thus, to zeroth order, the evolution with 
redshift of the neutral gas density matches that expected from gas depletion
due to star formation (eg. Lanzetta et al. 1991, Lanzetta, Wolfe \& Turnshek
1995, Storrie-Lombardi, McMahon \& Irwin 1996). Further, the evolution 
of metallicity with redshift also roughly matches what one would expect from
models of galactic evolution (Ferrini, Moll\'{a} \& D\'{i}az 1997, 
Fall 1997). 

	On the other hand, the morphology of damped~\lya systems remains
poorly understood. Based on the edge-leading asymmetries seen in the
absorption profiles of low ionization metals associated with these
systems, Prochaska \& Wolfe (1997) suggest that they are rapidly
rotating large disks with significant vertical scale heights. However, 
such profiles can also be explained by models in which damped systems are
much smaller objects that are undergoing infall and merger \cite{hsr}. It\ has 
also been claimed that the metal abundance of damped~\lya systems depends 
on the total HI column density in a way as would be expected from large
disks with central HI holes \cite{wp}, although the number of systems involved 
in this study is small. 

	For damped~\lya systems that lie in front of radio loud quasars, 
it is possible to augment the optical/UV spectra with HI 21cm absorption
spectra. Such a comparison, yields, among other things (and under suitable 
assumptions), the spin temperature, $T_s$, of the HI gas. Derived spin 
temperatures of damped~\lya systems have, in general, been much larger than
those observed in the disk of the Galaxy or in nearby galaxies 
(Braun \& Walterbos 1992, Braun 1996), implying that either damped~\lya 
systems are not disks, or that the ISM in the damped~\lya proto-disks is
considerably different from that in the local $z=0$ disks, presumably 
due to evolutionary effects.

	 Studies of low redshift damped~\lya systems are particularly 
interesting in this regard, since evolutionary effects are expected to
be negligible. Further, much more detailed information is obtainable, in 
particular from HST and/or ground based imaging, which makes identification
of the absorber possible (eg. Le Brun et al. 1997, Lanzetta et al. 1997).
Of course, it remains a possibility that the population that gives rise 
to damped~\lya absorption at low redshift is distinct from that at high
redshift. 

	In this paper, we report the detection of redshifted 21cm absorption
in two low redshift ($z= 0.2212$, $z=0.0912$) damped~\lya systems seen towards
the  quasar OI~363 (0738+313, $z_{em} = 0.630$) \cite{rt}. Observations of 
the higher redshift
system confirm, at considerably improved spectral resolution and 
sensitivity, earlier results from the Westerbork Synthesis Radio Telescope 
(WSRT) (Lane et al. 1998), while the lower redshift ($z=0.0912$) system is 
the lowest redshift damped~\lya system known to date. 

\section{Observations and Data Reduction}

	The observations were carried out using the GMRT (Swarup et al. 1991,
Swarup et al. 1997). 
The backend used was the proto-type eight station FX correlator, which 
gives a fixed number (128) of spectral channels over a total bandwidth 
that can be varied from 64~kHz to~16 MHz. Due to various ongoing maintenance
and installation activities, the actual number of antennas that were 
available during our observing runs varied between six and eight.

	For the observations of the  $z=0.0912$ system the bandwidth was
set to 1.0~MHz. No spectral taper was applied, giving a channel spacing 
of $\sim 1.8$~km s$^{-1}$.  Two observing runs were made, one on 27 June 1998 
and the other on 5 July 1998. The on source time for each run was about
six hours. Two observing runs were also taken for the  $z=0.2212$ system
(on 26 June 1998 and 4 July 1998), the first with a total bandwidth of
1.0~MHz (i.e. a channel spacing of $\sim 2.0$~km s$^{-1}$) and the other with 
a total bandwidth of 0.5~MHz (i.e. a channel spacing of $\sim 1.0$~km s$^{-1}$
 ). Each of these observing runs had an on source time of $\sim 4$~hours. 
Bandpass calibration at both redshifts was done using 3C~295, which was 
observed at least once during each observing run. 

	The data was converted from the raw telescope format to FITS and
then reduced in AIPS in the standard way. Maps were produced after 
subtracting out the continuum emission of the background quasar using UVLIN,
and spectra extracted from the resulting three dimensional cube. The GMRT
does not do online doppler tracking; this is, however, unimportant since the
doppler shift within any one of our observing runs was a small fraction of 
a channel. For the lower redshift system, data from the observations on 
different days were corrected to the heliocentric frame and then combined. 

\begin{figure}
\vskip -2.25cm
\epsfxsize=6.0cm 
\epsfysize=11.0cm 
\hskip -6.5cm \epsfbox{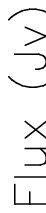}
\caption{GMRT redshifted 21cm absorption spectrum of the lower redshift 
         system towards OI363. The channel spacing is $\sim 1.8$~km s$^{-1}$.
         The deepest optical depth ($\sim 0.18$) is at a heliocentric 
         redshift of 0.09118. The width (FWHM) of the line is 
         $\sim$ 5~km s$^{-1}$.}
\label{fig:lz}
\end{figure}

	The final spectrum for the $z=0.0912$ system is shown in 
Figure~\ref{fig:lz}. There appear to be two components, one considerably 
deeper than the other. The fainter component, although weak, was detected 
in both our observing runs, and its magnitude is also considerably higher 
than the noise level. It is, of course, possible that the spectrum consists of
two components, one of which is broad and weak and the other, much deeper 
but narrow. The redshift of the narrow component is consistent with the 
redshift quoted in Rao \& Turnshek~(1998). The peak optical depth is 
$\sim 0.18$ (i.e. a depth of $390$~mJy with the continuum flux of OI363 
being 2.0~Jy), and occurs at a redshift of $z=0.09118 \pm 0.00001$. The 
FWHM of the line is small, $\sim$ 5~km s$^{-1}$.

\begin{figure}
\vskip -2.25cm
\epsfxsize= 6.0cm 
\epsfysize=11.0cm
\hskip -6.5cm \epsfbox{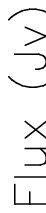}
\caption{GMRT redshifted 21cm absorption spectrum of the higher redshift
         system towards OI363. The channel spacing is $\sim 2.0$~km s$^{-1}$.
         The peak optical depth ($\sim 0.07$) occurs at a heliocentric 
         redshift of 0.2212. The width (FWHM) of the line is 
         $\sim$ 5.5~km s$^{-1}$}
\label{fig:hz}
\end{figure}

	Lane et al. (1998) report a redshift of $z=0.2212$ for the higher 
redshift system, based on WSRT observations. The $2.0$~km s$^{-1}$ GMRT 
spectrum (which has a considerably better velocity resolution and 
sensitivity than the WSRT spectrum) is shown in Figure~\ref{fig:hz}. The
redshift measured from this spectrum is $0.2212 \pm 0.00001$. This is
consistent with the redshift measured from the $1.0$~km s$^{-1}$ resolution
spectrum (which is not shown here). The peak optical depth ($\sim 0.07$)
is somewhat less than that of the lower redshift system, but the velocity
width is comparable, $\sim$ 5.5~km s$^{-1}$ (FWHM).

\section{Discussion}

	The total HI column density of a damped system can be determined
from its \lya profile; this can be then used, in combination with the measured 
HI 21cm optical depth, to determine the spin temperature of the absorbing 
gas (under the assumption that the absorbing gas is homogeneous. For a 
multi-phase absorber, this derived temperature is the column density weighted
harmonic mean of the spin temperatures of the different phases, provided
all the phases are optically thin.). One of the principal uncertainties in
this derivation of the spin temperature is the covering factor of the 
absorbing gas. In particular, the radio emission from quasars is often 
extended, while the UV continuum source is essentially a point source; 
thus, the line of sight along which the HI column density has been
derived from UV measurements need not be the same as the one for which 
the 21cm optical depth has been measured. The case of OI~363, however, is
relatively straightforward. OI~363 is a core dominated Gigahertz peaked
source whose total flux decreases from 2.2~Jy at 1.64~GHz to 1.59~Jy at
408~MHz. VLA maps at 1.64~GHz \cite{murphy} and 1.4~GHz \cite{ajit} 
show that the source is highly core dominated, with about 97\% of the total
flux in an unresolved core component. VLBI measurements at 5~GHz 
\cite{vlbi} show that the core size is $\sim 10$~milli arcseconds 
($\sim 16$~pc and $\sim 31$~pc at redshifts of 0.0912 and 0.2212 
respectively, for $H_0$ = 75~km s$^{-1}$ Mpc$^{-1}$ and $q_0 =0.5$). 
Stanghellini et al. (1997) note that their 5~GHz VLBI map recovers only 
77\% of the total flux measured by the VLA at 5~GHz. While it is unclear 
whether there is much change in the source size between 5~GHz
and 1.3~GHz, we know from IPS measurements \cite{sjk} that the upper limit 
on the core size at 330~MHz is 50~milli arcseconds. At both redshifts, 
the depth of the line (see Figures~\ref{fig:lz}~\&~\ref{fig:hz}) considerably 
exceeds the flux in the weak lobes, this implies that the absorbers 
must cover the central core. Given the small size of this central core, 
the covering factor is likely to be close to unity.

	The HI column densities inferred from the present observations,
in terms of the spin temperatures of the two damped systems, are $1.82
\pm 0.02 \times 10^{18} T_{s}$~atoms cm$^{-2}$ and $0.71 \pm 0.04 \times 
10^{18} T_{s}$~atoms cm$^{-2}$, for the lower and higher redshift systems, 
respectively. The column densities measured by Rao \& Turnshek (1998), 
from the damped Lyman-$\alpha$ lines, are $7.9 \pm 1.4 \times 10^{20}$~atoms 
cm$^{-2}$ and $1.5 \pm 0.2 \times 10^{21}$~atoms cm$^{-2}$, again in order 
of decreasing redshift. The spin temperatures obtained are hence 
$825 \pm 110$~K (for the $z=0.0912$ absorber) and $1120 \pm 200$~K 
(for the $z=0.2212$ system). For the higher redshift system, our measurement
agrees within the errors with that of Lane et al. (1998). The overwhelming
source of the (formal) uncertainty is in the determination of the HI column
density from the UV measurements. Thus, even at redshifts where no evolution
is expected, the derived spin temperature is significantly higher than that
typically seen in the Galaxy. If one assumes that the HI 21cm spectral width
is entirely due to thermal motions, the required kinetic temperatures are
$\sim 625$~K and $\sim 750$~K for the lower and higher redshifted system 
respectively, i.e. comparable to the derived spin temperatures. Note 
however, that in the ISM of the Galaxy, there is no stable neutral phase 
with temperature $\sim 1000$~K. On the other hand, such high spin 
temperatures appear common at both high and intermediate redshifts (see eg.
de Bruyn, O'Dea \& Baum 1996, Carilli et al. 1996, Lane et al. 1996,
Kanekar \& Chengalur 1997, Boiss\'{e} et al. 1998).  

	Ground based imaging of the OI~363 field \cite{rt} shows that
there are no spiral galaxies at small impact parameters to the line of
sight, contrary to the canonical model where damped systems arise in extended 
disks. Similarly, the next lowest redshift damped~\lya absorber (0850+4400,
Lanzetta et al. 1997) appears to be associated with an S0 galaxy, while, at
intermediate redshifts, the absorbers appear to be associated with galaxies
spanning a wide range of morphological types \cite{leBrun}. Interestingly,
at lower redshifts still, where imaging of HI 21cm emission is possible,
21cm absorption from quasar galaxy pairs appears to be associated more 
with tidal tails or other extended features of gas rich galaxies \cite{chris},
and not directly with the disks of large spirals. 
	
	While the low number density of damped Lyman-$\alpha$ systems 
at $z < 1$ makes it {\it a priori} extremely unlikely that two such systems 
might be found along the same line of sight, the VLBI map of OI~363 appears to
rule out the possibility of this line of sight being biased due to 
gravitational lensing. The current observations (and the absence of 
detectable gravitational lensing) do not however place strong constraints
on the surface density or mass of the absorbing systems.

	In summary, it appears that even at the lowest redshifts, gas outside
the disks of spiral galaxies and with apparent physical parameters 
considerably different from the ISM of nearby galaxies has a non-trivial 
contribution to the total absorption cross-section. This is consistent
with observations that, even for intermediate redshift damped~\lya absorbers,
the metallicity is considerably lower than typical solar values \cite{boisse}. 
Finally, the present GMRT observations also suggest that evolutionary 
effects may not play an important role in understanding why the derived 
spin temperature for damped~\lya systems are in general higher than those 
measured in nearby spiral galaxy disks. 

{\bf Acknowledgments} These observations would not have been possible
without the many years of dedicated effort put in by the GMRT staff to
build the telescope. The GMRT 1400~MHz wide-band feed and receiver system
was built by the Raman Research Institute. We are also grateful to Wendy 
Lane, Judith Irwin, Anish Roshi, D. J. Saikia, R. Srianand and Kandu 
Subramanian for their comments and suggestions.

\pagebreak 
\noindent {\bf Figure 1.} GMRT redshifted 21cm absorption spectrum of the
lower redshift system towards OI363. The channel spacing is $\sim 1.8$~km s$^{-1}$.
The deepest optical depth ($\sim 0.18$) is at a heliocentric redshift of
0.09118. The width (FWHM) of the line is $\sim$ 5~km s$^{-1}$.\\

\noindent {\bf Figure 2.} GMRT redshifted 21cm absorption spectrum of the higher redshift
system towards OI363. The channel spacing is $\sim 2.0$~km s$^{-1}$. The peak optical
depth ($\sim 0.07$) occurs at a heliocentric redshift of 0.2212. The width
(FWHM) of the line is $\sim$ 5.5~km s$^{-1}$.\\
\vskip 0.2 in
\bsp 

\label{lastpage}

\end{document}